# Dielectric and Magnetic Properties of (Bi$_{1-x}$La$_x$FeO$_3$)$_{0.5}$(PbTiO$_3$)$_{0.5}$ Ceramics Prepared from Mechanically Synthesized Powders

E. Markiewicz[a,*], B. Andrzejewski[a], B. Hilczer[a], M. Balcerzak[b], A. Pietraszko[c] and M. Jurczyk[b]

[a]Institute of Molecular Physics, Polish Academy of Sciences
Smoluchowskiego 17, PL-60179 Poznań, Poland
*corresponding author: ewamar@ifmpan.poznan.pl; tel.: +48 61 8695146; fax: +48 61 8684524

[b]Institute of Material Science and Engineering, Poznan University of Technology,
Piotrowo 3, 60-965 Poznań, Poland

[c]Institute of Low Temperatures an Structure Research, Polish Academy of Sciences,
Okólna 2, 50-422 Wrocław, Poland

*Abstract* — **Multiferroic (Bi$_{1-x}$La$_x$FeO$_3$)$_{0.5}$(PbTiO$_3$)$_{0.5}$ ceramics with $x$=0.0, 0.1, 0.2 and 0.5 was prepared from mechanical synthesized nanopowders. The XRD studies revealed the tetragonal structure and the tetragonality decreased with La content. Dielectric response of the compounds was found to contain three anomalies: 1) relaxor-like behavior due to lattice disorder (below 300 K); 2) dielectric permittivity maxima at ~400 K attributed to the presence of oxygen vacancies; 3) grain boundary effect above 475 K. The Curie point at ~500 K was observed for the compound with $x$=0.5. The composition near the morphotropic boundary: (Bi$_{0.8}$La$_{0.2}$FeO$_3$)$_{0.5}$(PbTiO$_3$)$_{0.5}$ shoved the highest remnant magnetization. The irreversible magnetic properties of the (Bi$_{1-x}$La$_x$FeO$_3$)$_{0.5}$(PbTiO$_3$)$_{0.5}$ compounds can be explained in terms of disorder induced spin-glass behavior due to random substitution of La or Pb ions for Bi sites. A sharp step in magnetization about 250 K is caused by the A-site distortion associated with tilts of FeO$_6$ octahedra leading to modification of Fe-O-Fe angles and of antiferromagnetic coupling between magnetic Fe$^{3+}$ moments.**

*Keywords-component; (Bi$_{1-x}$La$_x$FeO$_3$)$_{0.5}$(PbTiO$_3$)$_{0.5}$, dielectric properties, magnetic properties, relaxor-like behavior, lattice disorder*

## I. INTRODUCTION

Multiferroics as the materials which exhibit coupled ferroelectric, ferromagnetic and ferroelastic orderings have attracted considerable attention due to their potential for application in transducers, resistive switching elements, information storage, and spintronics [1-5]. Bismuth ferrite BiFeO$_3$ (BFO) a single phase multiferroic at room temperature, with antiferromagnetic-paramagnetic transition temperature $T_N$=640 K and ferroelectric Curie point at $T_C$=1100 K, is one of the most extensively studied materials [6] but the range of its application is restricted because of numerous drawbacks. The synthesis of pure BFO is very difficult due to comparable thermodynamic stability of Fe$^{2+}$ and Fe$^{3+}$ ions in the end compound [7] as well as the volatility of Bi$_2$O$_3$ starting oxide. Preparation from the oxides weighted in Bi$_2$O$_3$:Fe$_2$O$_3$ stoichiometric ratio leads to the appearance of undesired Bi$_2$Fe$_4$O$_9$ phase [8], whereas, the excess of Bi$_2$O$_3$ gives rise to the formation of parasitic Bi$_{25}$FeO$_{40}$ phase [8,9]. Moreover, according to the Bi$_2$O$_3$-Fe$_2$O$_3$ phase diagram [10], the temperature range of crystallization of BFO is very narrow: bellow the Curie temperature of 1098 K and above the eutectic point at 1058 K. In relevance to the fact, rapid heating and quenching following the conventional sintering should be involved in the synthesis process [8,11]. Another disadvantage of pure BFO is very weak magnetoelectric coupling, which is typical for multiferroics characterized by large difference between the transition temperatures $T_C$ and $T_N$. Theoretical studies by Gong and Jiang [12] pointed out the critical influence of spin-pair correlation fluctuation on the magnetodielectric effect which is the largest only around the magnetic phase transition temperature $T_N$. Kamba et al. [13] found that relative changes in the dielectric permittivity induced by magnetic field up to 9 T at the temperature above 250 K and bellow 200 K, *i.e.*, in the range where the Maxwell - Wagner mechanism does not contribute to the permittivity, are very low (~10$^{-5}$). At room temperature, the magnetic properties of BFO are determined by the structure of $G$ – type with a cycloidal spiral arrangement of the magnetic moments of Fe$^{3+}$ ions [14-16] and the canted spins due to the Dzialoshinskii–Moriya interaction [17] resulting in a weak magnetoelectric effect [18]. In addition to the mentioned drawbacks, BFO shows semiconducting behavior above 300 K and its ferroelectric hysteresis loop is well pronounced only at the temperatures below 100 K [13,19].

The synthesis of BFO - based multiferroics with high polarization and strong coupling between polarization and magnetization still remains a challenge. Many research laboratories make efforts to bring the ferroelectric $T_C$ and antiferromagnetic $T_N$ transition temperatures closer to each other. Generally applied method is the chemical modification by introduction other $ABO_3$ compounds to form solid solutions with BFO. $PbTiO_3$ (PT) has been considered to be the most promising candidate in this field of application by Venevtsev et al. in 1960 [20]. The most interesting results were reported by Zhu, Guo and Ye who established both the structural and the magnetic phase diagram of $(1-x)BiFeO_3-xPbTiO_3$ system [8]. The authors revealed the existence of rhombohedral phase ($x \leq 0.20$), orthorhombic phase ($0.20 \leq x \leq 0.28$) and tetragonal phase with large tetragonality ($x \geq 0.31$). Moreover, the authors considered the $(1-x)BiFeO_3-xPbTiO_3$ solution as a magnetically diluted system from $BiFeO_3$, stated that in the tetragonal phase the lattice parameter $c$ determines the "critical coupling distance" of the cooperative magnetic interaction of the magnetic ordering and established the magnetic phase diagram of the system. They stressed that the compositions with $BiFeO_3:PbTiO_3$ ratio around 50:50 favors a formation of chemically ordered microregions (with superlattices of $Fe^{3+}$ and $Ti^{4+}$), which can be considered as magnetic nanoclusters. Freezing of the magnetic nanoclusters results in a residual magnetization (a weak ferromagnetic state) at low temperatures.

Another method to improve the magnetoelectric effect in $BiFeO_3$ was initialized by Pająk, Połomska and Kaczmarek [21,22] and later by Sosnowska et al. [23]. Neutron diffraction studies performed on $Bi_xLa_{1-x}FeO_3$ (BLFO) single crystal showed that the incorporation of $La^{3+}$ ions at $A$ – site of the perovskite lattice also destroys the space modulated spin structure of BFO. A decade later the interest on La- doped $(1-x)BiFeO_3-xPbTiO_3$ system has been revived and numerous papers have been published [24-31]. Cheng et al. [24] reported a downward shift of the ferroelectric Curie temperature and a decrease in the tetragonality with increasing La content in $0.45BiFeO_3-0.55PbTiO_3$. Moreover, the material was found to exhibit magnetic ordering. Also the switching behavior of $La^{3+}$ doped $(1-x)(BiFeO_3)-x(PbTiO_3)$ solid solutions ($0.35 \leq x \leq 0.46$) was found to be sensitive to the c/a value [27]. The properties of $La^{3+}$ doped $(BiFeO_3)_{0.5}(PbTiO_3)_{0.5}$ with lanthanum content $0 \leq x \leq 0.5$ synthesized by solid-state reaction technique were studied by Singh et al. and by Mishra et al. [26,28-30,32]. Mishra et al. studied X-ray diffraction of $(Bi_{1-x}La_xFeO_3)_{0.5}(PbTiO_3)_{0.5}$ ceramics with average grain sizes of the order of 1 μm and reported the $T$-$x$ phase diagram for $x=0$, 0.2 and 0.3 with a downward shift of the tetragonal-cubic transition temperature [30]. A significant enhancement of magnetic and ferroelectric properties as well as the magnetoelectric coupling coefficient was claimed for $(Bi_{0.5}La_{0.5}FeO_3)_{0.5}(PbTiO_3)_{0.5}$ ceramics with maximum grain size of ~7 μm and cubic symmetry ($Pm3m$ space group) at room temperature [26]. The paper as well as the papers devoted to Raman spectra of $(Bi_{1-x}La_xFeO_3)_{0.5}(PbTiO_3)_{0.5}$ [28,29] were however, very confusing since the authors reported ferroelectric properties and Raman scattering in the cubic symmetry. In next paper [32] Singh et al. basing on powder neutron and X-ray diffraction data established the room temperature structure of $(Bi_{0.5}La_{0.5}FeO_3)_{0.5}(PbTiO_3)_{0.5}$ as rhombohedral with space group $R3c$ and reported high values of real and imaginary parts of dielectric permittivity in the temperature range 483 K$\leq T \leq$573 K. The enhancement of the dielectric properties has been related by the authors to space charge polarization due to piling up electric charges onto the interfaces of ceramic grains and grain boundaries. Nevertheless, interesting dielectric relaxor-like behavior have been reported in the temperature range 100÷300 K for highly doped $(Bi_{1-x}La_xFeO_3)_{0.5}(PbTiO_3)_{0.5}$ ceramics with $x=0.5$ [26,28].

Magnetic properties of $(Bi_{1-x}La_xFeO_3)_{1-y}(PbTiO_3)_y$ were for the first time reported by Cheng et al. in the case of solid solutions with $y=0.55$ and $x=0.1$, 0.2 and 0.3 [26]. The authors observed symmetric magnetic hysteresis loops, characteristic of magnetically ordered materials, with magnetization increasing with $La^{3+}$ doping. Mishra et al. presented non-saturated magnetic hysteresis loops for $(Bi_{1-x}La_xFeO_3)_{0.5}(PbTiO_3)_{0.5}$ solid solutions with $x=0÷0.5$ at low temperatures [29]. The most interesting were however, their results on temperature variation of the magnetization measured in field cooling (FC) and zero field cooling (ZFC) conditions, which point to a spin-glass behavior of the solid solution with $x=0.2$. The spin-glass properties have not been observed for $x \geq 0.3$ and following the considerations given by Zhu et al. [8] the authors related the behavior to the $Fe^{3+}$-O-$Fe^{3+}$ along the $c$-direction in the tetragonal phase. A serious discussion on the role of the covalence of A-O and B-O bonds in the perovskite structure of $(Bi_{1-x}La_xFeO_3)_{0.6}(PbTiO_3)_{0.4}$ in the ferroic state formation was given recently by Cótica et al. [33].

After considering the mentioned above drawbacks of pure BFO as well as the suggested methods of improving the magnetoelectric coupling we looked for a single phase multiferroic solid solution the properties of which can be tuned by aliovalent doping of bismuth ferrite compounds with ferroelectric perovskites. As the compounds of BFO with $PbTiO_3$ exhibit structural change from rhombohedral to tetragonal structure [8] and $La^{3+}$ doping may turn the structure again to the rhombohedral symmetry [6, 23] we proceeded to study the dielectric and magnetic response of $(Bi_{1-x}La_xFeO_3)_{0.5}(PbTiO_3)_{0.5}$ ceramics with $x=0$, 0.1, 0.2 0.3 and 0.5. The ceramics was obtained from powders prepared by mechanosynthesis from respective oxides and was characterized by X-ray diffraction (XRD) and scanning electron microscopy (SEM).

## II. EXPERIMENTAL

*2.1 Preparation of the samples*

The synthesis of BLFO-PT samples with various content of La was carried out in SPEX 8000 Mixer Mill for 48 h at room temperature in the air. The milling time was optimized using XRD pattern for various times. The starting oxides $Bi_2O_3$ (99,975% Alfa Aesar), $La_2O_3$ (99,99% Aldrich), $Fe_2O_3$ (99% Aldrich), PbO (99,9% Aldrich) and $TiO_2$ ($\geq$99% Aldrich) were weighted in stoichiometric ratios. The weight ratio of the stainless steel balls to the oxides was 2:1. The as-milled powders were heat treated at 700°C for 0.5 h in the air and pressed into pellets at the pressure of 800 MPa. For dielectric measurements the samples ~1 mm thick and 8 mm in diameter were covered with gold sputtered electrodes in Baltec SCD 050 sputter coater.

*2.2 XRD studies*

The structure of the synthesized BFO was controlled by room-temperature X-ray powder diffraction studies using X'Pert-PANalytical diffractometer with CuK$_\alpha$ radiation. The data were collected at room temperature in $2\theta$ range from $20^0$ to $80^0$ after milling and after the heat-treatment of the sample. The mean crystallite sizes were calculated from half-width of the most intensive peak (110), by using Scherrer equation and Panalytical High Score program.

To obtain more precise information on the crystallographic structure of $(Bi_{0.5}La_{0.5}FeO_3)_{0.5}(PbTiO_3)_{0.5}$ solid solution the diffraction patterns were collected at the Bragg–Brentano geometry in the reflection mode in the range of $10°≤2\theta≤95°$ during 18 h at room temperature with a scan step of 0.026 deg. An X'Pert PRO X-ray diffraction system with CuK$_\alpha$ radiation equipped with a PIXcel ultra-fast line detector, focusing mirror and Soller slits was used. The crystal structure of the solid solution was refined using the Rietveld method in a program package High Scor Plus. The model of the crystal structure was taken from the ICSD base and the refinement was performed in the standard setting the *R3c* space group and the *P4mm* space group.

*2.3 SEM studies*

Scanning electron micrographs were obtained at room temperature using digital microscope VEGA TS5135 Tescan with the resolution of 3.5 nm. The incident electron energy amounted to 20 keV.

*2.4 Dielectric measurements*

Dielectric response of the ceramic samples was studied using an Alpha-A High Performance Frequency Analyzer (Novocontrol GmbH) combined with Quatro Cryosystem for the temperature control. The sample with gold-evaporated electrodes was fixed between two additional external electrodes in a sample holder and placed into a cryostat. The measurements were performed in the temperature range from 125 K to 575 K on heating at a rate of 1 K/min. The frequency varied from 1 Hz to 1 MHz at the oscillation voltage of 1 V. The measured dielectric permittivity data were collected and evaluated by WinDETA impedance analysis software and a WinFit V 3.2. program.

*2.5 Magnetic measurements*

Magnetic measurements were performed using a Vibrating Sample Magnetometer probe installed on a Quantum Design Physical Property Measurement system (PPMS) fitted with a superconducting 9 T magnet. Magnetization loops were registered up to a maximum field of ±5 T at ambient temperature and down to 4 K. The zero field cooled (ZFC) and field cooled (FC) magnetization were measured as a function of temperature in the temperature range 4÷300 K.

## III. RESULTS AND DISCUSSION

*3.1. Structural studies*

Fig. 1 shows X-ray diffraction patterns of $(Bi_{1-x}La_xFeO_3)_{0.5}(PbTiO_3)_{0.5}$ powders with various concentration x of La ions after 48 hours of milling respective oxides and annealing at 973 K for 0.5. The X–ray diffraction patterns show sharp peaks which in Fig. 2 are indexed according to the *P4mm* structure. The XRD study does not reveal the presence of any parasitic phase although the primarily performed EDS measurement showed a trace of Ni atoms.

The average crystallite sizes calculated from half-width of the most intensive peak (110), are presented in Table 1. Mean sizes of the crystallites varies from 15 to 28 nm increasing with increasing La contents. These crystallites agglomerate to bigger particles with size from several to around 50 μm (Fig. 2a). After pressing at 800 MPa powder material creates bulk material with a porosity of 16.6% (Fig. 2b).

Table 1. Crystallite size calculated for $(Bi_{1-x}La_xFeO_3)_{0.5}(PbTiO_3)_{0.5}$ with various La contents *x*.

| Lanthanium content *x* | 0 | 0.1 | 0.2 | 0.3 | 0.5 |
|---|---|---|---|---|---|
| Crystallite size (nm) | 15 | 19 | 26 | 27 | 28 |

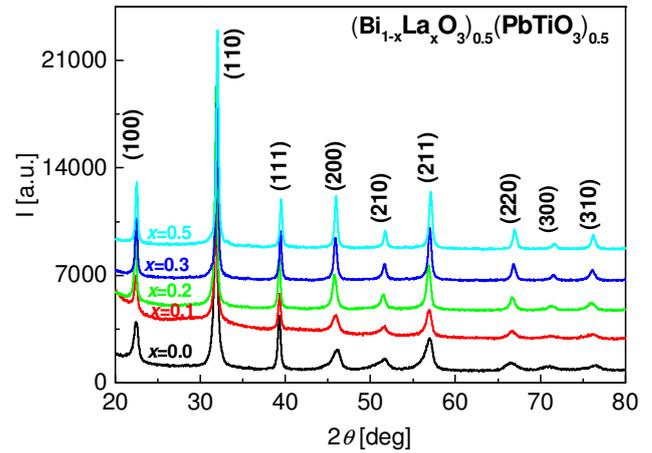

Figure 1. XRD patterns of $(Bi_{1-x}La_xFeO_3)_{0.5}(PbTiO_3)_{0.5}$ powders after 48 hours of milling of the respective oxides and heat treatment at 973 K for 0.5 h; (Indexing for tetragonal symmetry).

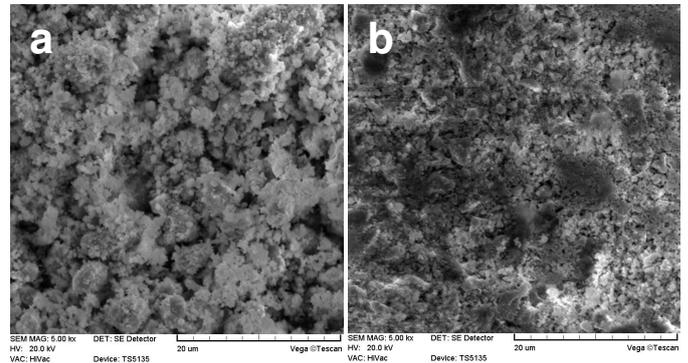

Figure 2. SEM picture of mechanically milled and additionally annealed at 973 K for 0.5h $(Bi_{0.8}La_{0.2}FeO_3)_{0.5}(PbTiO_3)_{0.5}$ powder (a) and bulk material (pressed at 800 MPa) (b).

The powder X-ray diffraction data of $(Bi_{1-x}La_xFeO_3)_{0.5}$ $(PbTiO_3)_{0.5}$ ceramics with La- content $x$=0, 0.1 and 0.2 were refined using Rietveld method and a PANalytical program to the tetragonal structure with the *4Pmm* (No 99) space group. The dependences of the lattice parameters on the La contents are presented in Fig. 3. One can observe the decrease of tetragonality as the concentration of $La^{3+}$ ions increases. The result is in a good agreement with the literature data [24,27,29,30].

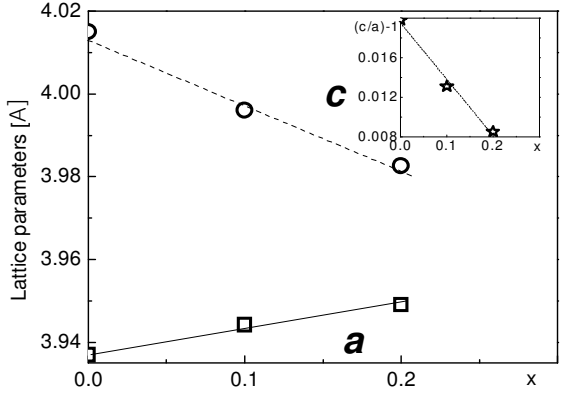

Figure 3. Dependence of lattice parameters of $(Bi_{1-x}La_xFeO_3)_{0.5}(PbTiO_3)_{0.5}$ ceramics on La contents $x$; the inset shows a decrease in the tetragonality with $x$.

The problem appears with the structure of $(Bi_{1-x}La_xFeO_3)_{0.5}$ $(PbTiO_3)_{0.5}$ solid solution of the highest lanthanum content $x$=0.5 which was found to have rhombohedral symmetry [32]. To resolve the structure of our $(Bi_{0.5}La_{0.5}FeO_3)_{0.5}(PbTiO_3)_{0.5}$ nanoceramics we proceeded to analyze higher order reflections from the X-ray diffraction data collected during 18 h in the Bragg-Brentano geometry. Fig. 4 shows the X-ray pattern of the ceramics, whereas the analysis of the X-ray reflections of higher order is presented in Fig. 5. The experimental X-ray high reflection profiles (thick red line) are compared with those obtained by simulation for a standard setting of the rhombohedral structure with *R3c* space group (No 161) and the profiles simulated for a tetragonal symmetry *P4mm* space group (No 99) from the ICSD base.

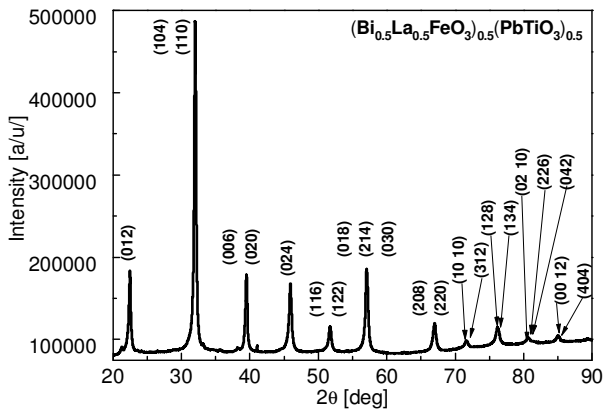

Figure 4. Dependence X-ray diffraction pattern of $(Bi_{0.5}La_{0.5}FeO_3)_{0.5}$ $(PbTiO_3)_{0.5}$ ceramics collected during 18 h in the reflection mode and Bragg-Brentano geometry.

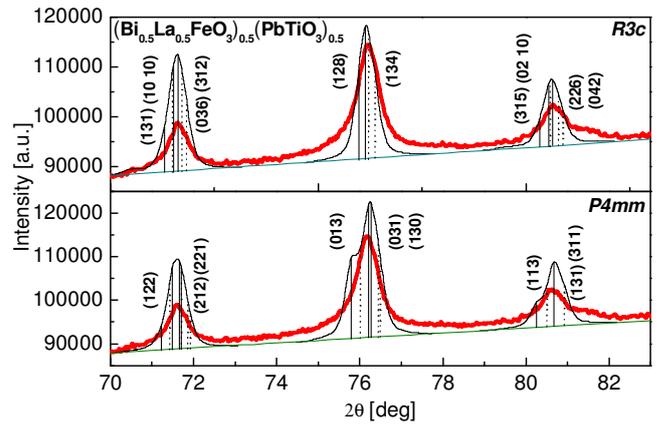

Figure 5. Higher order X-ray reflections obtained during 18 h in the reflection mode and Bragg-Brentano geometry for $(Bi_{0.5}La_{0.5}FeO_3)_{0.5}$ $(PbTiO_3)_{0.5}$ ceramics (thick red line); thin lines are the result of simulation of the reflections in the rhombohedral (*R3c*) and tetragonal structure (*P4mm*).

One can observe that the shape of X-ray high reflection profiles speaks in favor of the rhombohedral *R3c* structure. The lattice parameters obtained from the Rietveld refinement and the $R_{Bragg}$ factors are presented in Table 2. It should be however, observed that the results are on the limit of our X-ray experiment and therefore we analyzed the case of the highest La - doping only.

Table 2. Lattice parameters and $R_{Bragg}$ factors obtained from Rietveld refinement of the X-ray diffraction pattern of $(Bi_{0.5}La_{0.5}FeO_3)_{0.5}(PbTiO_3)_{0.5}$ obtained in the reflection mode and Bragg-Brentano geometry.

| Space group | a [Å] | b [Å] | c [Å] | α=β [deg] | γ [deg] | $R_{Bragg}$ % |
|---|---|---|---|---|---|---|
| *R3c* | 5.5825(4) | 5.5825(4) | 13.740(3) | 90 | 120 | 3.84647 |
| *P4mm* | 3.941(3) | 3.941(3) | 3.9675(4) | 90 | 90 | 3.74202 |

*3.2 Dielectric response*

Fig. 6 shows temperature dependences of the real part of dielectric permittivity $\varepsilon'(f,T)$ at frequencies $f$=100 Hz, 158 Hz, 251 Hz, 398 Hz, 631 Hz, 1 kHz, …10 kHz, …100 kHz, … 631 kHz, 1 MHz obtained for $(Bi_{1-x}La_xFeO_3)_{0.5}(PbTiO_3)_{0.5}$ ceramics of various La content $x$. Two anomalies can be identified for the samples with La contents $x \leq 0.3$. The low-temperatures dielectric anomaly, *i.e.* below 300 K, is more apparent in the dielectric absorption representation $\varepsilon''(f,T)$ and will be discussed later. In the vicinity of 400 K at lower frequencies one can observe a trace of dielectric anomaly that increases with the La- doping and at $x \geq 0.3$ transforms into a frequency dependent anomaly. We would like to ascribe this anomaly to the presence of oxygen vacancies [34]. For the $(Bi_{1-x}La_xFeO_3)_{0.5}(PbTiO_3)_{0.5}$ ceramics with La content $x$=0.5 one can observe the Curie point anomaly at ~500 K. More details relevant to the effect of La content on the dielectric properties of ceramics provides the analysis of dielectric absorption spectra presented in Fig. 7.

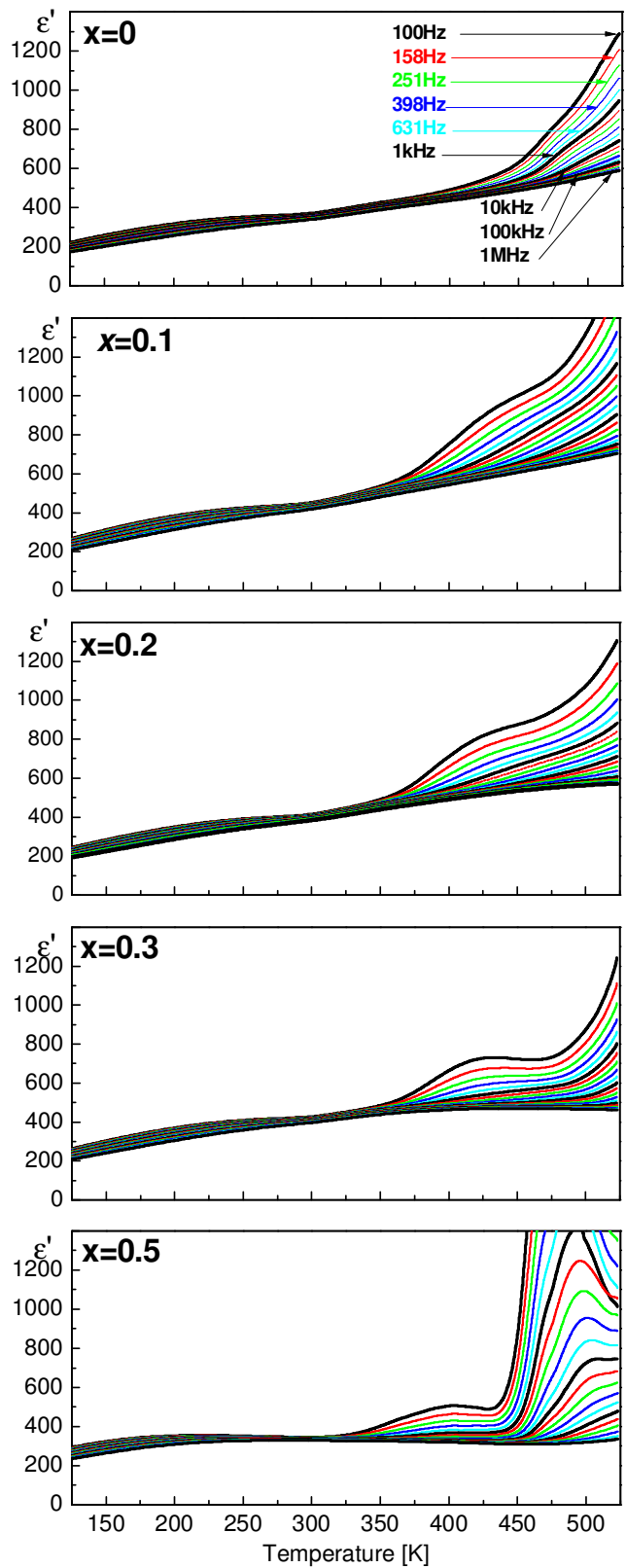
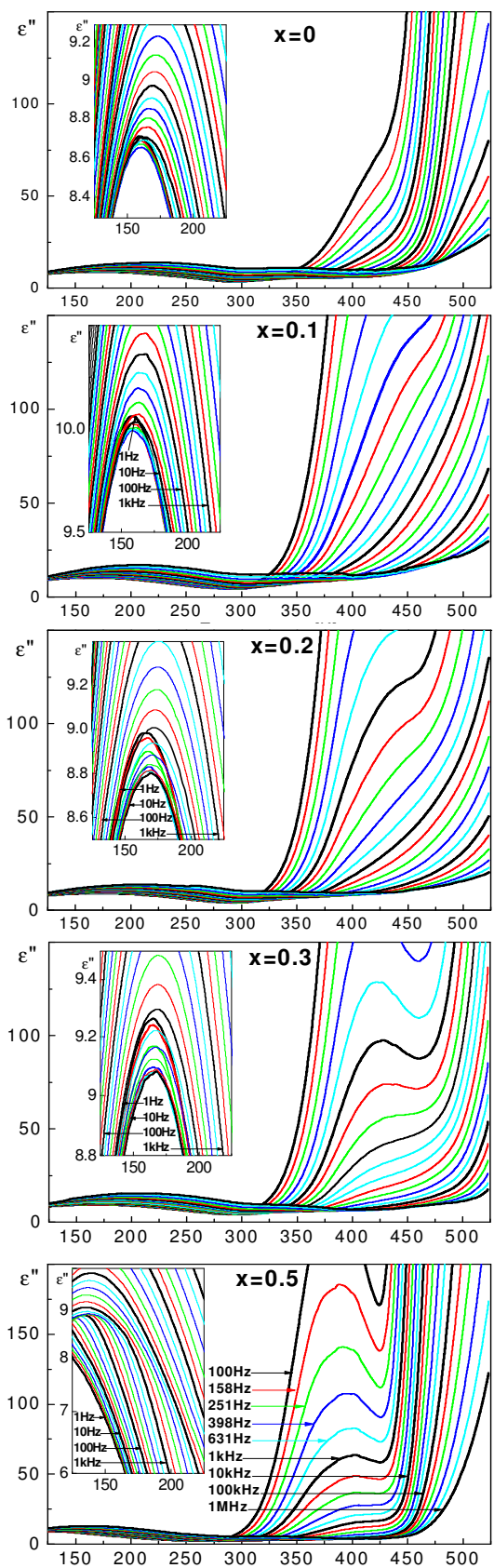

Figure 6. Temperature dependences of real part of dielectric permittivity $\varepsilon'$ obtained for $(Bi_{1-x}La_xFeO_3)_{0.5}(PbTiO_3)_{0.5}$ with various contents $x$ of $La^{3+}$ ions.

Figure 7. Temperature dependence of imaginary part of dielectric permittivity $\varepsilon''$ obtained for $(Bi_{1-x}La_xFeO_3)_{0.5}(PbTiO_3)_{0.5}$ with various contents $x$ of $La^{3+}$ ions.

The anomaly observed below 300 K for all investigated samples points to relaxor–like behavior that is a characteristic of perovskite compounds based solid solution and results from a compositional disorder. The incorporation of $La^{3+}$ and $Pb^{2+}$ cations at the $Bi^{3+}$ site of the BFO lattice and incorporation $Ti^{4+}$ at the $Fe^{3+}$ site creates local random electric and stress fields due the difference in the valences and ionic radii. In highly polarizable matrix the compositional disorder results in formation of polar nanoregions (PNRs) [34-36]. The dielectric response of compositionally disordered solid solution of perovskite type is determined by the response of dipole moments flipping between allowed directions in PNRs and by fluctuation of polar PNR boundaries. The contributions has been considered in details by Bokov and Ye [37,38]. The contribution originating from fluctuations of nanoregion boundaries is apparent at low-frequencies as a decrease in dielectric losses maxima $\varepsilon''_{max}$ with increasing frequency $f$, whereas reorientations of the dipoles within the PNRs result in an increase of $\varepsilon''_{max}$ values with increasing frequency. The contributions can be seen in the insets of Fig. 7 and the frequency range in which the particular behavior is dominant is given in Table 3. The relaxor-like behavior observed in $(BiFeO_3)_{0.5}(PbTiO_3)_{0.5}$ solid solution is due the difference in the valences and ionic radii both in the A and B- sites in the lattice (in the A- site: between $Bi^{3+}$ and $Pb^{2+}$, in the B site: between $Fe^{3+}$ and $Ti^{4+}$) and the maximum of the dielectric absorption at 1 MHz appears at ~218 K. Additional doping with $La^{3+}$ ions causes a downward shift of the $\varepsilon''(f,T)$ anomaly characteristic of the relaxor state: $T_{\varepsilon''max}$ at 1 MHz appears at ~209 K, ~208 K and ~206 K for x =0.1, 0.2 and 0.3, respectively. For highly doped $(Bi_{0.5}La_{0.5}FeO_3)_{0.5}(PbTiO_3)_{0.5}$ the low-frequency part lies below the temperature range available, whereas the maximum absorption at 1 MHz appears at ~160 K. The huge shift in $T_{\varepsilon''max}$ for $(Bi_{0.5}La_{0.5}FeO_3)_{0.5}(PbTiO_3)_{0.5}$ with respect to that for undoped BFO-PT ceramics may be due to the change in the crystal structure. It should be also mentioned that due to overlapping with the dielectric losses related to oxygen vacancies relaxation we are not able to determine the temperature range of the relaxor-like response.

Table 3. Frequency ranges in which contributions to the dielectric response from PNR boundary fluctuations and dipole moments reorientations in PNRs were observed by us for $(Bi_{1-x}La_xFeO_3)_{0.5}(PbTiO_3)_{0.5}$ solid solution.

| Main contributions to the dielectric response | x=0 | x=0.1 | x=0.2 | x=0.3 |
|---|---|---|---|---|
| Fluctuation of PNRs boundaries $\varepsilon''_{max}$ decreases with increasing frequency | 1Hz÷4Hz | 1Hz÷4Hz | 1Hz÷10Hz | 1 Hz÷16Hz |
| Reorientations of dipole moments in PNRs $\varepsilon''_{max}$ increases with increasing frequency | 6.3Hz÷1MHz | 6.3Hz÷1MHz | 16Hz÷1MHz | 25Hz÷1MHz |

The relaxation times of the low temperature process follow the Vogel–Fulcher relationship:

$$\tau = \tau_0 \exp(E_{VF}/(T-T_0)). \quad (1)$$

where $E_{VF}$ denotes the activation energy, $T_0$ stands for the temperature of freezing of dipolar motions in PNRs and $\tau_0$ is the pre-exponential factor. Fig. 8 shows the relaxation times determined for all investigated samples versus reciprocal temperature. The values of activation energy $E_a$ and freezing temperature $T_0$ resulting from fitting to the formula (1) are summarized in Table 4. One can observe a significant decrease in the activation energy due to doping with lanthanum.

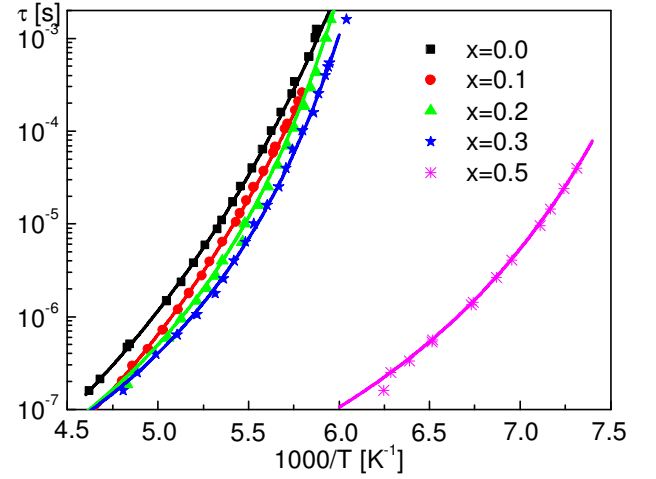

Figure 8. Relaxation times of dielectric anomaly related to the structural disorder versus reciprocal temperature determined for $(Bi_{1-x}La_xFeO_3)_{0.5}(PbTiO_3)_{0.5}$ samples with La contents $x$=0, 0.1, 0.2, 0.3 and 0.5 The symbols are the experimental data and the lines are the Vogel-Fulcher plots.

Table 4. Activation energy $E_{VF}$ and freezing temperature $T_0$ obtained from fitting temperature dependences of the relaxation times to the Vogel–Fulcher formula (1) for $(Bi_{1-x}La_xFeO_3)_{0.5}(PbTiO_3)_{0.5}$ ceramics.

| $x$ | $E_{VF}$ [K] | $T_0$ [K] |
|---|---|---|
| 0.0 | 1006.7 (0.087 eV) | 116.6 |
| 0.1 | 491.3 (0.042 eV) | 137.0 |
| 0.2 | 453.4 (0.039 eV) | 143.7 |
| 0.3 | 436.6 (0.038 eV) | 137.3 |
| 0.5 | 361.7 (0.031 eV) | 106.5 |

To characterize the process of high-temperature electrical transport in $(Bi_{1-x}La_xFeO_3)_{0.5}(PbTiO_3)_{0.5}$ solid solution ac conductivity ($\sigma_{ac}$) versus frequency was measured in the temperature range from 470 K to 525 K with the temperature step of 5 K. Fig. 9 shows an example of frequency dependence of the electric conductivity at various temperatures for $(Bi_{1-x}La_xFeO_3)_{0.5}(PbTiO_3)_{0.5}$ solid solution with $x$=0.2. One can observe that at low frequencies, the conductivity $\sigma_{ac}$ saturates to a constant value *i.e.* to the dc conductivity $\sigma_{dc}$.

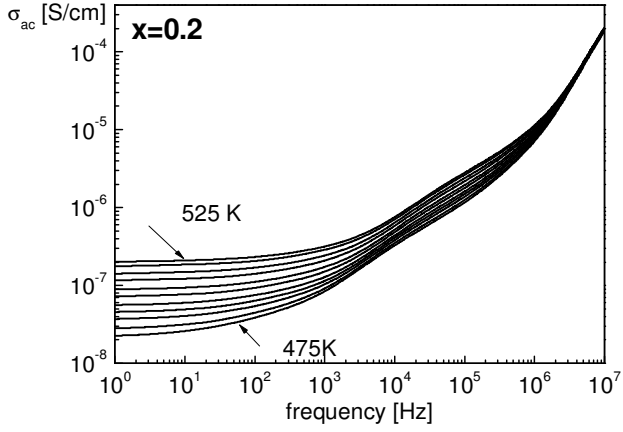

Figure 9. Frequency dependence of ac conductivity $\sigma_{ac}$ at constant temperatures from the range 475÷525 K for $(Bi_{0.8}La_{0.2}FeO_3)_{0.5}(PbTiO_3)_{0.5}$ ceramics.

In Fig. 10, the $\sigma_{dc}$ conductivity obtained for $(Bi_{1-x}La_xFeO_3)_{0.5}(PbTiO_3)_{0.5}$ samples are plotted as the function of reciprocal temperature. One should notice that the plots obey the Arrhenius law:

$$\sigma_{dc} = \sigma_0 \exp(E_a / k_B T). \qquad (2)$$

where $\sigma_0$ is the pre-exponential factor and $E_a$ denotes the activation energy for electrical transport. The calculated values of activation energy $E_a$ and the pre-exponential factors are given in Table 5.

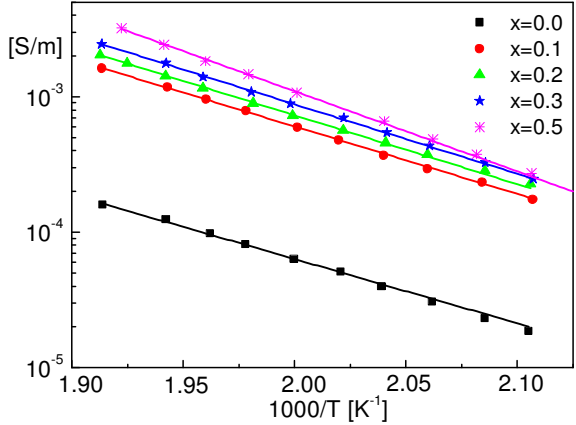

Figure 10. Frequency dependence of ac conductivity $\sigma_{ac}$ at constant temperatures from the range 475÷525 K for $(Bi_{0.8}La_{0.2}FeO_3)_{0.5}(PbTiO_3)_{0.5}$ ceramics.

Table 5. Activation energy $E_a$ and pre-exponential factor $\sigma_0$ according to the Arrhenius law for $(Bi_{1-x}La_xFeO_3)_{0.5}(PbTiO_3)_{0.5}$ samples with various La contents $x$.

| $x$ | $E_a$ [eV] | $\sigma_0$ [S/m] |
|---|---|---|
| 0.0 | 0.94 | $2.0 \cdot 10^5$ |
| 0.1 | 0.99 | $5.8 \cdot 10^6$ |
| 0.2 | 1.01 | $1.1 \cdot 10^7$ |
| 0.3 | 1.02 | $1.6 \cdot 10^7$ |
| 0.5 | 1.17 | $7.6 \cdot 10^8$ |

One can see that the La doping of $(BiFeO_3)_{0.5}(PbTiO_3)_{0.5}$ causes an increase in the activation energy responsible for migration of oxygen vacancies. It should be noted that the values of $E_a$ close to 1 eV are typical for thermally activated motion of double-ionized oxygen vacancies as reported for some perovskite materials like $Bi:SrTiO_3$ (1.09 eV) [39], $SrTiO_3$ (0.98 eV) [40] and $YFeO_3$ (1.08 eV) [41].

### 3.3 Magnetic properties

Fig. 11 shows the magnetization versus magnetic field $M(H)$ measured for $(BiFeO_3)_{0.5}(PbTiO_3)_{0.5}$ and $(Bi_{1-x}La_xFeO_3)_{0.5}(PbTiO_3)_{0.5}$ with $x=0.2$ at constant temperatures of 300, 200, 100 and 4 K. The $M(H)$ curves for BFO-PT were found to exhibit linear dependence in a wide range of magnetic fields and temperatures, which indicates an antiferromagnetic ordering. A trace of nonlinear behavior in $M(H)$ appears at low temperatures for lanthanum doped BLFO-PT samples because of enhanced spin canting and ferromagnetic ordering. Also, one cannot exclude some paramagnetic contribution to the total magnetization originating from contaminations in highly doped BLFO-PT or from disordered BFO lattice. One can come to a conclusion that the incorporation of $La^{3+}$ ions at $A$-site of the BFO structure destroys the space modulated spin ordering and enhances the Dzyaloshinskii–Moriya interaction between $Fe^{3+}$ ions resulting in canted spin arrangement of unpaired electrons. The remnant magnetization $M_r$ was found to be dependent on the La content and reached the highest value in the vicinity of the morphotropic boundary (Table 6).

Table 6. Remnant magnetization $M_r$ at $T=4$ K for $(Bi_{1-x}La_xFeO_3)_{0.5}(PbTiO_3)_{0.5}$ solid solutions with various La contents.

| $x$ | 0 | 0.1 | 0.2 | 0.3 | 0.5 |
|---|---|---|---|---|---|
| $M_r$ [Am$^2$/kg] | 0.007 | 0.014 | 0.047 | 0.023 | 0.021 |

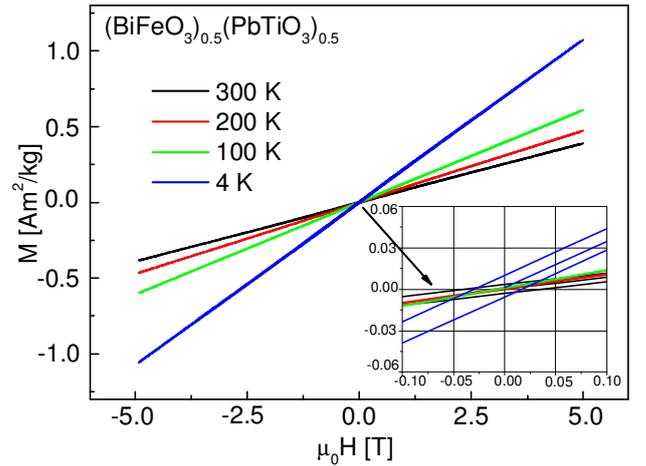

Figure 11. (a) Magnetization $M(H)$ versus magnetic field measured for $(BiFeO_3)_{0.5}(PbTiO_3)_{0.5}$ at temperatures of 300, 200, 100 and 4 K.

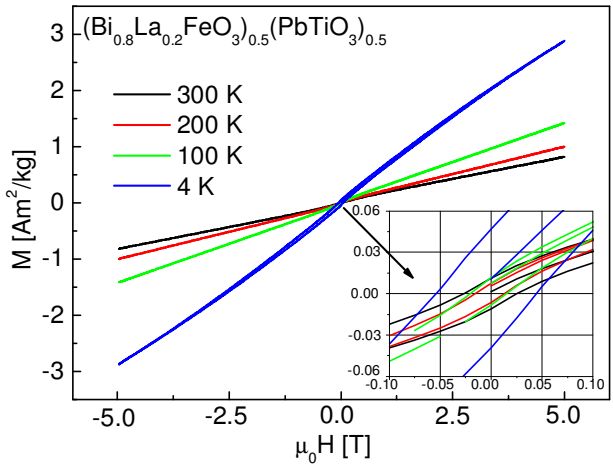

Figure 11. (b) Magnetization $M(H)$ versus magnetic field measured for $(Bi_{0.8}La_{0.2}FeO_3)_{0.5}(PbTiO_3)_{0.5}$ at temperatures of 300, 200, 100 and 4 K.

Magnetic properties studies revealed in the temperature range 4÷300 K two contributions to the magnetization of the $(Bi_{1-x}La_xFeO_3)_{0.5}(PbTiO_3)_{0.5}$ nanoceramics pointing to the coexistence of weak ferromagnetic clusters and antiferromagnetic or paramagnetic ordering. The first contribution is apparent in a form of magnetic hysteresis loop, whereas the second one is responsible for linear variation of the magnetization $M$ with the magnetic field. Indeed, the weak ferromagnetic ordering in bismuth ferrite due to Dzyaloshinskii-Moriya type interaction was theoretically predicted by Ederer et al. [18] using local density spin approximation (LSDA) method. Later, weak ferromagnetic domains with the mean size about 30 nm corresponding to a half-period of the spin cycloid have been found experimentally in BFO by means of neutron diffraction [43].

Temperature variation of the magnetization $M(T)$ was measured in zero field cooling (ZFC) conditions and on cooling in a constant magnetic field of 0.1 T (FC condtions). Fig. 12 shows temperature dependences of the magnetization for $(Bi_{1-x}La_xFeO_3)_{0.5}(PbTiO_3)_{0.5}$ samples with $x=0$ (a), 0.1 (b) and $x=0.3$ (c). The magnetization was observed to decrease with temperature and to exhibit the effect of irreversibility. The temperatures $T_{irr}$ at which the magnetization determined for both FC and ZFC branches merges was found to be shifted upwards with increasing La content (Table 7). The effect is similar to that observed by Mishra at al. [30].

Table 7. The temperature of irreversibility $T_{irr}$ for $(Bi_{1-x}La_xFeO_3)_{0.5}(PbTiO_3)_{0.5}$ solid solution with various La contents.

| $x$ | 0 | 0.1 | 0.2 | 0.3 | 0.5 |
|---|---|---|---|---|---|
| $T_{irr}$ [K] | 222.8 | 262.3 | 264.9 | 265.4 | 266.5 |

At low temperatures, about 10 K one can observe a trace of a cusp in ZFC magnetization for samples with $x=0$ and 0.1. This feature is attributed to a spin glass behavior resulting from the randomness and magnetic frustration due to a competition between ferromagnetic and antiferromagnetic orders. Similar behavior has been reported for other perovskite-type materials in which the nonmagnetic ions share the B site, e.g. $Pb(Fe_{1/2}Ta_{1/2})O_3$ [44] or $La_2CoMnO_6$ [45]. The disappearance of the cusp due to spin glass behavior for $x=0.2$ and also for $x=0.3$ evidences the significant role of La-doping in decreasing the temperature of magnetic ordering due to lower tetragonality which facilitates cooperative magnetic interaction.

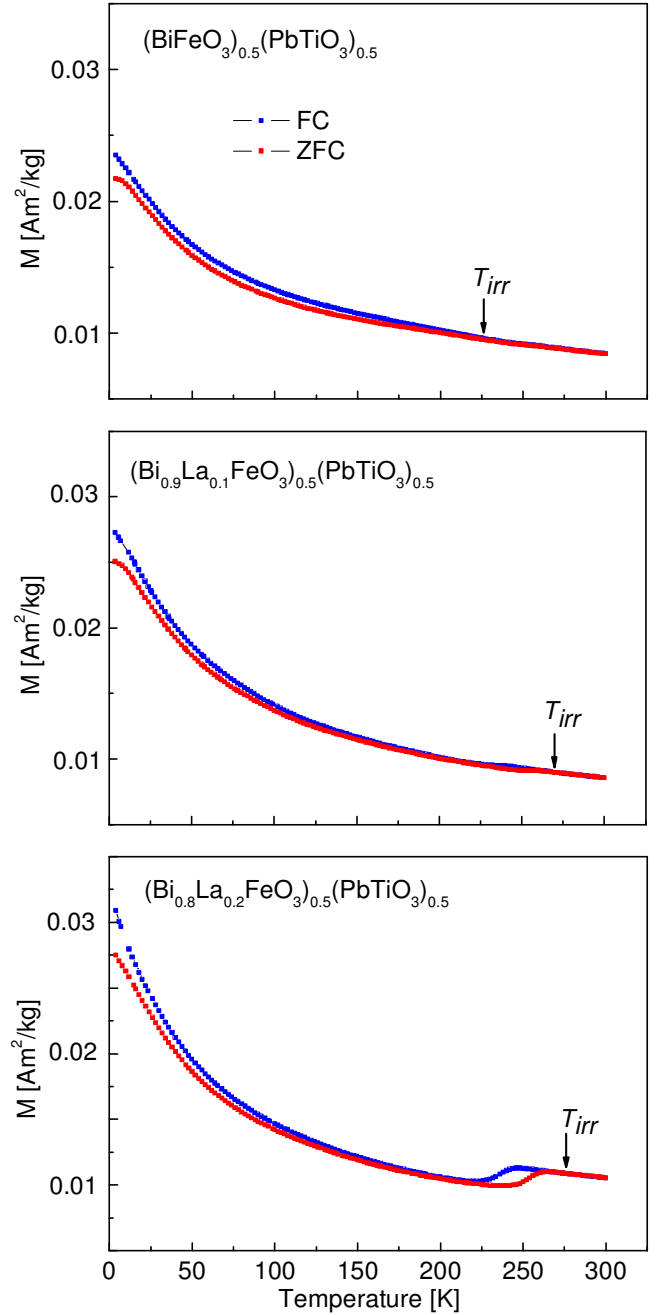

Figure 12. Temperature dependences of magnetization $M(T)$ for $(Bi_{1-x}La_xFeO_3)_{0.5}(PbTiO_3)_{0.5}$ with $x=0$ (a), 0.1 (b) and $x=0.3$ (c) in ZFC and FC ($\mu_0H$=0.1 T) conditions.

One can observe that the magnetization of $La^{3+}$ doped $(Bi_{0.7}La_{0.3}FeO_3)_{0.5}(PbTiO_3)_{0.5}$ solid solution exhibit an abrupt change by amount of about 10% near the temperature of 250 K (Fig. 12 c). It seems that this feature can be related to magnetically driven structural transitions found recently in $Bi_{0.5}La_{0.5}FeO_3$ compound by Kavanagh et al. [46]. The Rietveld analysis performed by this group shows that when the temperature approaches about 300 K, the $FeO_6$ octahedra become more regular with respect to the Fe-O bonds while the internal Fe-O-Fe angles start to deviate from the ideal tetrahedral geometry. This distortion correlates with A and O sites displacements represented by $R_5^+$[A site] and $R_5^+$[O site] modes. Both Fe-O1-Fe and Fe-O2-Fe angles (where O1 is the axial oxygen and O2 is the equatorial oxygen), depend on the $FeO_6$ block tilts and also on the orthorhombic distortion. The changes of temperature modify the $FeO_6$ octahedra tilts which can be represented by two possible modes: the in phase ($M_3^+$) and antiphase ($R_4^+$) rotations. It turns out that the antiphase rotation affects the Fe-O1-Fe angle whereas the Fe-O2-Fe angle is influenced by both in-phase and antiphase rotations. The exact values of Fe-O1-Fe and Fe-O2-Fe angles are determined by the superposition of these $FeO_6$ block tilts and of octahedral distortions. The tilts as well as the octahedral distortion are temperature dependent and cause a significant increase in the Fe-O1-Fe angle with simultaneous decrease in the Fe-O2-Fe angle when temperature rises from 50 K to about 300 K. In higher temperature both angles do not alter and again start to increase above about 700 K.

From the point of view of magnetic properties of oxides, the Fe-O1-Fe and Fe-O2-Fe angles are crucial parameters because they modify superexchange coupling and in this way determine magnetic ordering. The step in the FC and ZFC magnetizations observed near 250 K can be now consistently explained in terms of the superexchange coupling modification due to the changes of the Fe-O1-Fe and Fe-O2-Fe angles. It seems that this feature is mainly related to the reduction of the Fe-O2-Fe angle as the temperature increases, which results in a weakening of the antiferromagnetic coupling and in an increase of the total magnetization. The magnitude of the magnetization step depends on the lanthanum doping level i.e. it is almost negligible for undoped $(BiFeO_3)_{0.5}(PbTiO_3)_{0.5}$ compound and increases with amount of $La^{3+}$ doping reaching about 8% of the total magnetization for $(Bi_{0.5}La_{0.5}FeO_3)_{0.5}(PbTiO_3)_{0.5}$ compound. This behavior clearly demonstrates that $La^{3+}$ and $Bi^{3+}$ ions located in A-site of the bismuth ferrite structure are not equivalent because they displace along *a*-axis with various amplitudes [46], which causes anomalous orthorhombic distortion and in-phase tilt [46]. The modifications of the BLFO-PT compound structure, change Fe-O-Fe angles and thus strongly influence its magnetic ordering. Note also, that the most pronounced magnetic anomaly about 250 K occurs for $x=0.3$ i.e. close to the morphotropic boundary where the tetragonal and rhombohedral phases exist simultaneously. Therefore, it can be explained in alternative way as antiferromagnetic transition in tetragonal phase. Similar magnetic transitions were observed near 250 K by Zhu et al. [8] for $(1-x)BiFeO_3$-$xPbTiO_3$ solid solutions in morphotropic phase boundary region. On the other hand, random substitution of La for Bi at the A-sites of the BFO lattice introduces some disorder to the crystal structure. As the doping level increases the amount of disorder should increase which manifests in the enhanced temperature of irreversibility $T_{irr}$ between ZFC and FC magnetizations due to spin-glass behavior (see table 7).

## IV. CONCLUSIONS

A multiferroic ceramics (BLFO-PT) prepared from mechanically synthesized powders was proposed for applications in the electronic industry as well as its dielectric and magnetic properties were investigated. The experimental data confirm that the structural disorder caused by substitution of $La^{3+}$ and $Pb^{2+}$ cations at $A$ – site and $Ti^{4+}$ at $B$ – site in the perovskite lattice is very useful tool to control the shift in the Curie point. For the lanthanum content of 0.5 mol %, the position of $T_C$ was observed at ~500 K, i.e. about 140 K below the $T_N$ temperature. One can expect that the BLFO-PT composition with $x=0.2$ will enable the location of $T_C$ closer to $T_N$ and it would be in a special accordance with the best magnetic properties obtained for that composition. The measurements in the vicinity of $T_N$ will be the subject of our future study.

## V. ACKNOWLEDGEEMENTS

The XRD, SEM, AFM and dielectric response studies were supported by the funds from COST Action MP0904: Single- and multiphase ferroics and multiferroics with restricted geometries (SIMUFER). The magnetic studies were supported by National Science Centre in Poland (project N N507 229040).